# Co-propagation of QKD & 6 Tb/s (60 × 100G) DWDM Channels with ~17 dBm Total WDM Power in Single and Multi-Span Configurations

P. Gavignet, E. Pincemin, F. Herviou, Y. Loussouarn, F. Mondain, A. J. Grant, L. Johnson, R. I. Woodward, J. F. Dynes, B. Summers, A. J. Shields, K. Taira, H. Sato, R. Zink, V. Grempka, V. Castay, and J. Zou

***Abstract*—We report co-propagation experiments of the quantum channel (at 1310 nm) of a Quantum Key Distribution (QKD) system with Dense Wavelength Division Multiplexing (DWDM) data channels in the 1550 nm range. Two configurations are assessed. The first one is a single span configuration where various lengths of Standard Single Mode Fiber (SSMF) (from 20 to 70 km) are used and the total WDM channels power is varied. The Secure Key Rate (SKR) and the Quantum Bit Error Ratio (QBER) are recorded showing that up to ~17 dBm total power of 30 or 60 channels at 100 Gb/s can coexist with the quantum channel. A metric to evaluate the co-propagation efficiency is also proposed to better evaluate the ability of a QKD system to provide secure keys in a co-propagation regime. The second experiment is a three spans link with a cascade of three QKD systems and two trusted nodes in a 184 km total link length. We report the transmission of a coherent 400 Gb/s Dual Polarization DP-16QAM (Quadrature Amplitude Modulation) channel that transports a QKD secured 100 GbE data stream, with other fifty-four 100 Gb/s WDM channels. Encryption is demonstrated at the same time as co-propagation.**

## I. Introduction

THE security of the data that are transported in the networks of a telecommunication operator is one of its main concern as it forms the basis of mutual trust with its customers. People working on security activities have always been searching to implement the best security solutions in order to efficiently protect the data of the customers. However, for the past decade, the threat of quantum computers breaching today's security solutions is becoming more and more possible. Besides the field of Post-Quantum Cryptography (PQC), whose goal is to find quantum robust algorithms, a non-algorithmic, physics-based solution has been proposed which is called Quantum Key Distribution (QKD) [1]. Even if this solution was first proposed to improve the security of communication networks, it is also the first step of the future quantum internet.

Many solutions have been tested in laboratories, some of them in field trials but very few have been implemented in operational networks. A large deployment of this new technology by operators still faces some reluctance like the cost and constraints of introduction of QKD systems in the operators' networks. Indeed, when an operator wants to deploy QKD, it has to deal with various constraints of operational networks, such as the scarcity of the resource for a QKD-dedicated fiber infrastructure, the mismatch between transmission range of QKD and WDM systems, and the impossibility for quantum signals to pass through an optical amplifier.

Since several decades, network operators are increasing the transmission capacity of their infrastructure and, especially in Orange, we have succeeded in introducing new technologies while conserving the already deployed fibers. A prerequisite to the introduction of QKD at large scale would thus be the ability to maintain this approach. Moreover, the very high performance of the Wavelength Division Multiplexing (WDM) links either in terms of reach or capacity depend on the engineering rules of the deployed WDM systems and the introduction of QKD should be compatible with these rules. Finally, the cost of a dedicated fiber infrastructure for the quantum channel would be unaffordable. That is why the co-propagation, on the same fiber, of the quantum channel and existing WDM channels is of utmost importance. This co-propagation often reduces the Secure Key Rate (SKR) which is particularly sensitive to the length of the transmission link, through an increased Quantum Bit Error Ratio (QBER), principally due to spontaneous Raman scattering [2], [3] of co-propagating light.

Another important point when dealing with long haul networks is the possibility to transmit an end to end key over long distances (several hundreds of kilometers or more). With currently available QKD commercial equipment, and according

P. Gavignet, E. Pincemin, F. Herviou, and Y. Loussouarn are with Orange Innovation, 22300 Lannion, France (e-mail: paulette.gavignet@orange.com; erwan.pincemin@orange.com;fabrice.herviou@orange.com;y.loussouarn@orange.fr).

F. Mondain was with the Orange Innovation, 22300 Lannion, France. He is now with Exail, 75002 Paris, France (e-mail: fr.mondain@gmail.com).

A. J. Grant, L. Johnson, R. I. Woodward, J. F. Dynes, B. Summers, and A. J. Shields are with Toshiba Europe Ltd., CB4 0WW Cambridge, U.K. (e-mail:andy.grant@toshiba.eu;Lee.johnson@toshiba.eu;robert.woodward@toshiba.eu;james.dynes@toshiba.eu;benedict.summers@toshiba.eu; andrew.shields@toshiba.eu ).

K. Taira and H. Sato are with Toshiba Digital Solutions Corporation, Kawasaki 212-8585, Japan (e-mail: kentaro.taira@toshiba.co.jp ; hideaki1.sato@toshiba.co.jp).

R. Zink is with the Adva Network Security GmbH, 12489 Berlin, Germany (e-mail: richard.zink@advasecurity.com).

V. Grempka, V. Castay, and J. Zou are with the Adtran Networks SE, 91140 Villebon-sur-Yvette, France, and also with the Adtran Networks SE, 82152 Munich, Germany (e-mail: valentin.grempka@adtran.com; vincent.castay@adtran.com; jim.zou@adtran.com).

to the implemented technical solution, it is necessary to establish a quantum key on each span of a WDM multi-span link and relay a “global key” from end to end using a Quantum Key Management System (Q-KMS). This builds what is known as a “trusted node QKD network”.

In this article, we report the results of two experiments:

--The first one is dedicated to the classical & quantum co-existence on several tens of kilometers of Standard Single Mode Fiber (SSMF). Very high aggregated power (~17 dBm) of up to 60 x100 Gb/s DWDM channels is accepted by the QKD system and high SKRs are obtained which makes this configuration fully compatible with currently deployed WDM transmission links. A metric to evaluate the co-propagation efficiency is also proposed to ease the performance comparison of classical & QKD channels co-propagation experiments.

--The second experiment is performed on a three spans link whereby three QKD systems have been installed. This allows the encryption of a 100 GbE flow of a 400G channel. The 400 Gb/s channel under study is inserted into a 54 wavelengths 100 Gb/s DWDM comb with a total power up to ~17.5 dBm. This encrypted signal is transported over three QKD links of, respectively, 67 km, 50 km and 67 km of SSMF which are connected to each other via trusted nodes.

Compared to the work published in [4], this paper goes a step further with a multi-span configuration enabling to cascade several QKD systems and providing an OTNsec [5] (Optical Transport Network) QKD-based encryption over 184 km.

## II. Single Span Experiment

### A. Description of the multiplexed QKD system

For the first experiment, we use a Multiplexed (MU) QKD system, designed and built by Toshiba. This system implements an efficient BB84 QKD protocol with decoy states [6]. Phase-encoded quantum states are generated by a gain-switched laser followed by an asymmetric Mach-Zehnder interferometer in the QKD transmitter, and avalanche photodiodes (APDs) are used in the receiver to measure received states [7]. The total system size is 3 rack-units (3U) per node.

The QKD system is optimized for supporting co-propagation of quantum and classical signals. This is achieved using a quantum channel at 1310 nm, for increased spectral separation from the QKD service channel / multiplexed data channels in the telecom C-band, to minimize the impact of Raman scattering. In addition, the system employs high-extinction spectral filtering and time-domain gating at the QKD receiver to isolate the quantum channel from co-propagating / Raman-scattered light with maximal signal to noise ratio. The QKD system also includes automatic self-optimization routines to dynamically adjust various optical parameters for optimal performance on each communication link. Add/drop multiplexing hardware is also included in the QKD unit to multiplex the quantum channel, QKD service channels and any auxiliary data channels onto a fiber pair communication channel.

### B. Experimental evaluation of the co-propagation

The experimental set-up (Fig. 1) used for the co-propagation evaluation consists of a DWDM comb with sixty DP-QPSK (Dual Polarization Quadrature Phase Shift Keying) channels at 100 Gb/s ranged from 1533.6 nm to 1557 nm (with 50 GHz spacing) and sent to the Aux input (Rx) of the Alice side terminal. This optical comb is based on the interleaving of two amplified DWDM combs (with EDFAs (Erbium-Doped Fiber Amplifiers)) of 30 (100 GHz spaced) channels each. Another EDFA is placed after the interleaver leading to up to 18.5 dBm of WDM total input power at the Alice Aux Rx. In that case the WDM total power in the fiber (at the QKD Tx port) is 16.8 dBm due to an insertion loss of 1.7 dB on the path between Aux Rx and QKD Tx (see Fig. 1). A variable optical attenuator allows adjustment of the WDM aggregated power co-propagating with the quantum channel at 1310 nm. Various spools of SMF-28 single mode fiber are used (from 20 to 70 km) and a tap coupler (10/90) is inserted after the fiber to perform power or spectra measurements. In order to maximize the launch power for co-propagation, only the Alice to Bob direction is fed with the WDM comb. To further increase the power per channel considered, half of the comb can be removed. This will be indicated when appropriate. The spectra are also shown in Fig. 1 and we see that, in addition to the quantum channel (not on the spectra), the DWDM channels co-propagate with two QKD service channels (C59 & C60), which are responsible for performing the classical post-processing as part of the QKD protocol.

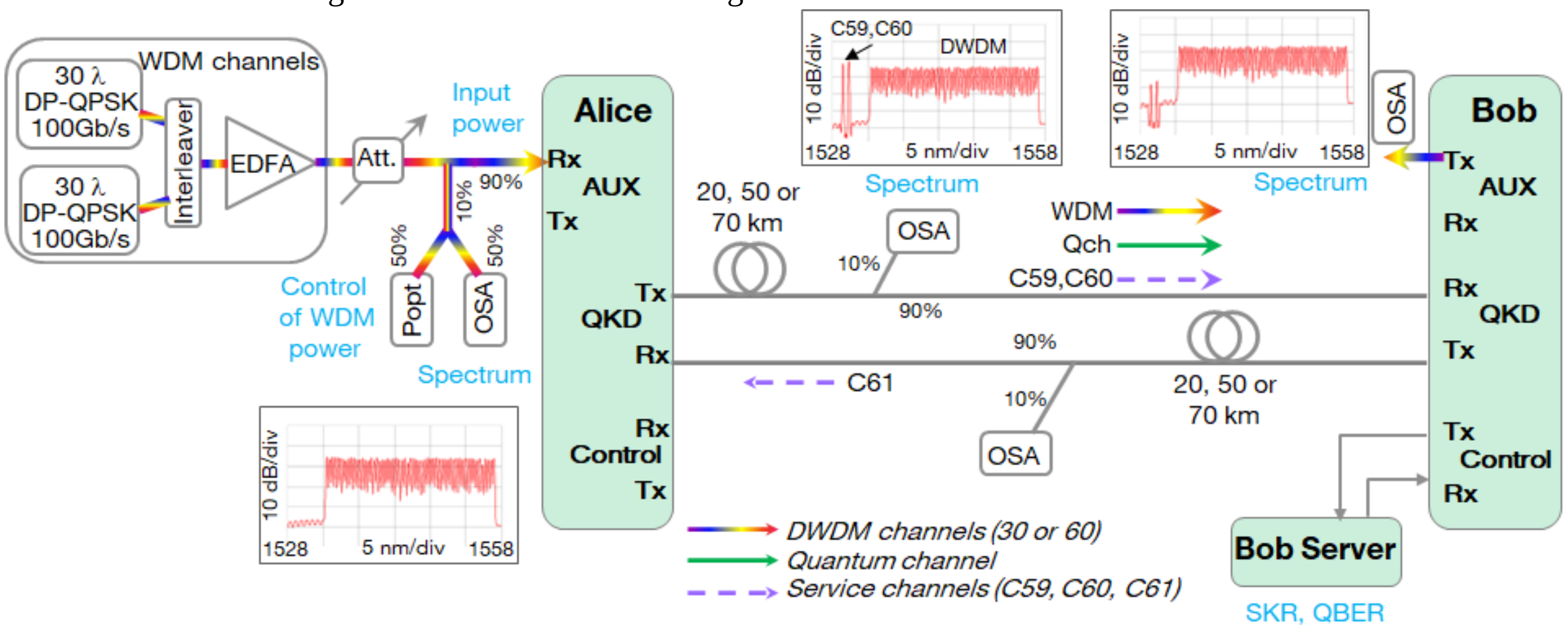

Fig. 1. Experimental set-up for co-propagation evaluation.

Fig. 2 shows the SKR (a) and the QBER (b) versus elapsed time for 50 km fiber length for a WDM total power of 16.8 dBm which is the maximum value allowed by the set-up. Green points correspond to the absence of WDM (wo: without WDM), blue triangles are with 30 channels and red squares are for 60 channels. The mean SKRs with 0, 30 or 60 channels are respectively 169, 107 and 106 kb/s. The corresponding mean QBERs are 3.4, 5.4 and 5.4 %.

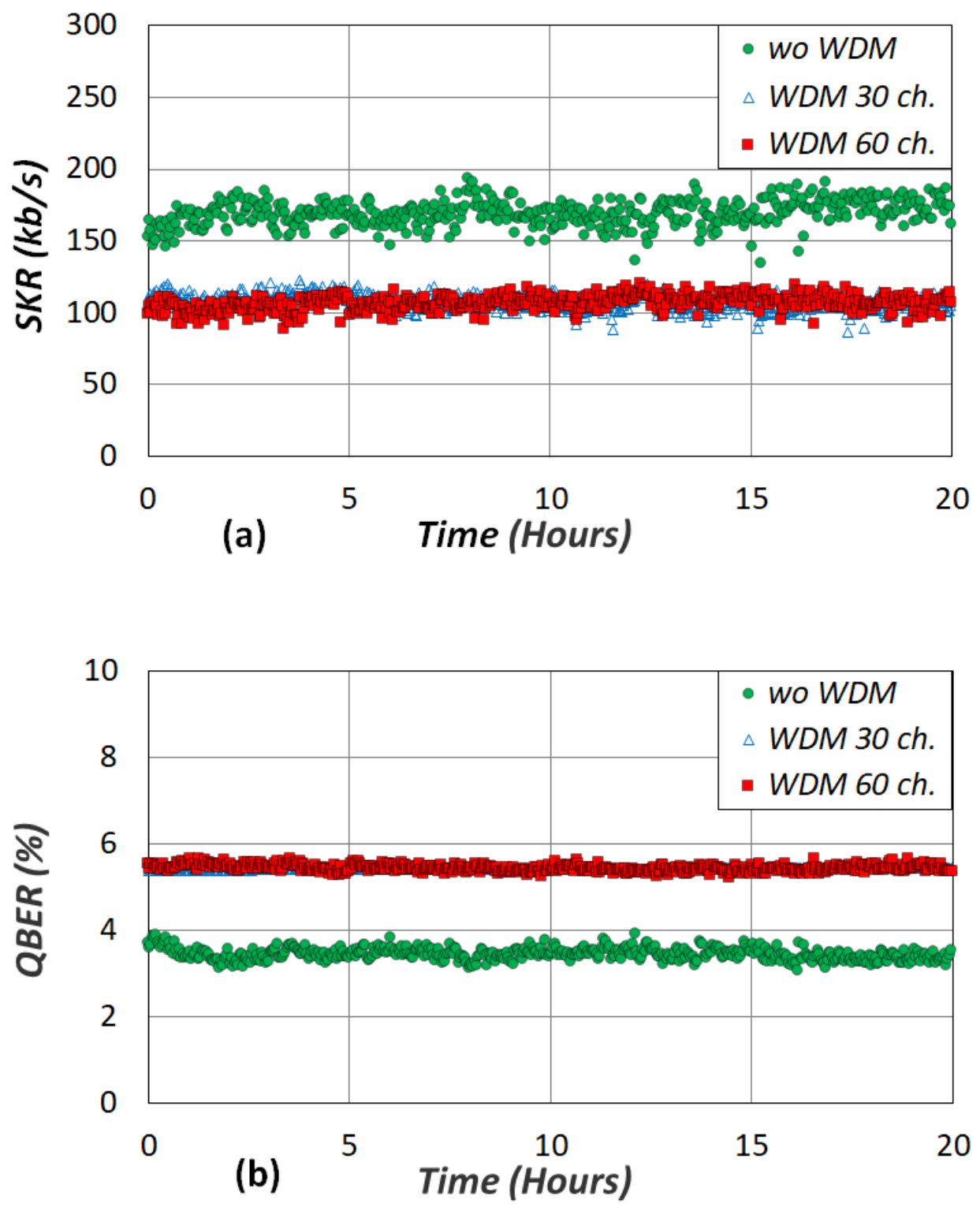

Fig. 2. SKR (a) and QBER (b) with 50 km fiber and 16.8 dBm WDM total power in the fiber. Power/channel ~ +2 dBm with 30 ch., ~ -1 dBm with 60 ch.

We also performed the tests with 20 and 70 km and Fig. 3 shows the evolution of the SKR versus fiber length for a constant WDM total power of 15.3 dBm in the fiber that ensures a fair comparison between the various fiber lengths under study and also allows to protect the QKD Rx from damages at 20 km (where the link losses are very low). For a fiber length of 20 km, the average SKR recorded for 64 hours is equal to 1.47 Mb/s with 60 channels and 15.3 dBm WDM total power. As previously mentioned the maximum WDM power available in the set-up did not allow us to reach the maximum tolerable WDM power at 50 km as it is higher than 16.8 dBm co-propagating in the fiber.

For 70 km we have reached the maximum acceptable power: it corresponds to 15.8 dBm total WDM power co-propagating with the quantum channel. It still allows to provide a SKR of 1.94 kb/s (average over 22 hours), which is a good performance, and sufficient for practical applications such as regularly rekeying of link encryptors. Beyond 15.8 dBm total WDM, the QKD system was not able to provide any key. Note that the total link loss with 70 km between Alice and Bob (from QKD Tx to QKD Rx) is 17.5 dB at 1550 nm and 25.7 dB at 1310 nm.

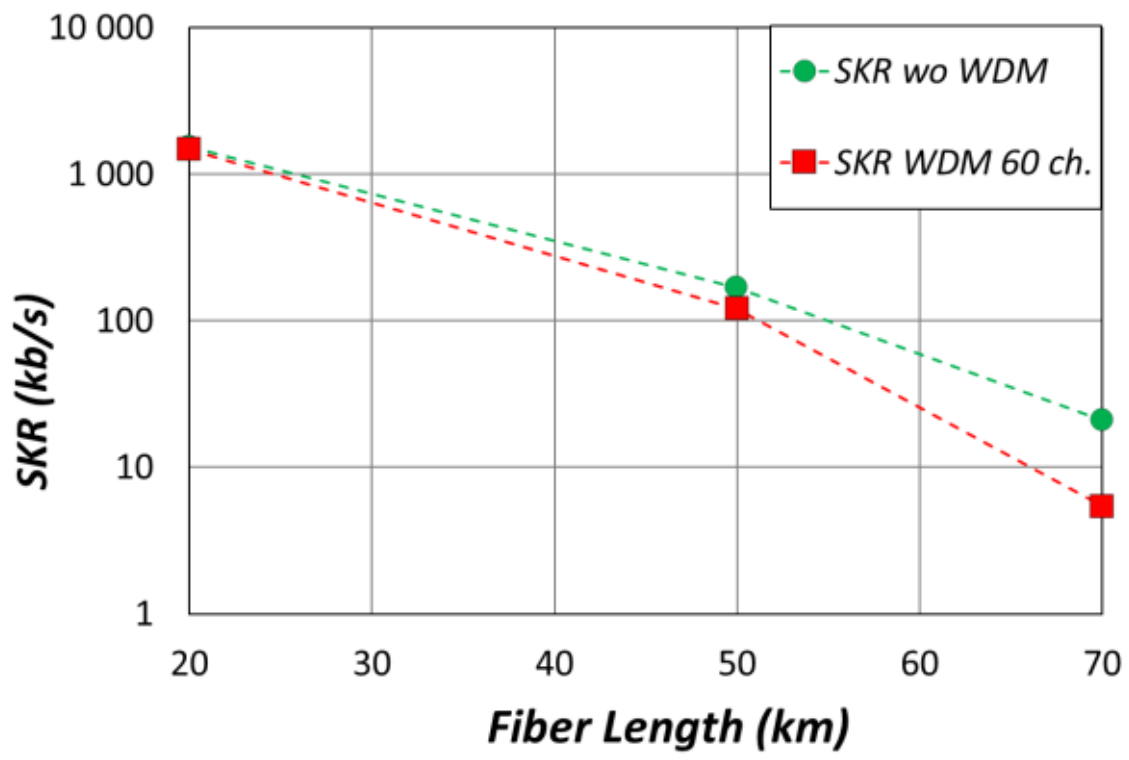

Fig. 3. SKR versus fiber length with 15.3 dBm WDM total power. Power/channel ~ +2 dBm with 30 ch., ~ -1 dBm with 60 ch.

### C. *Analysis of the results*

The main goal of these tests was to evaluate the possible deployment of QKD system on a WDM link that is already deployed in the field and determine the number of WDM channels that could be mixed with the QKD signal in the same SMF-28 fiber (to allow DWDM & QKD channels co-existence). If we consider the results of Fig. 2(a) with 50 km, we see that the SKR is higher than 100 kb/s and the results are the same in the two configurations (30 and 60 channels); this is also the case for the QBER (Fig. 2(b)). Moreover, we can see that the SKR and QBER values are very stable during time. We have presented results for a period of about 20 hours, but measurements have been performed for several days in various configurations and confirm the stability over time. The power per channel is ~ +2 dBm in the 30 channels configuration and ~ -1 dBm with 60 channels which is in the range of the power per channel currently used in DWDM's 100 Gb/s transmissions. In [8] it is shown that a power per channel of 0 dBm allows the transmission of 60 WDM channels at 100 Gb/s up to 1200 km. In [9] the transmission of a 400 Gb/s channel in a WDM comb with 57 other 100 Gb/s channels over 300 to 400 km with a power per channel in the range -2 to +5 dBm is possible. In both cases the results were obtained with pluggable devices making us confident in the co-propagation of several tens of WDM channels with the quantum channel in real field environment. We also performed tests feeding only the Bob to Alice direction with the DWDM channels and confirmed there was no influence on the SKR and QBER since the quantum signal travels on a separate fiber from the backward directed traffic.

### D. *Comparison with previous work - Introduction of new metric*

We also notice in Fig. 2 that the results in terms of SKR and QBER for a given fiber length only depend on the aggregated WDM power. This shows that the number of channels and/or the total bit rate that co-propagate with the quantum channel do not constitute the best parameters to consider, when evaluating the ability of co-existence of classical & QKD channels. Indeed, these experiments confirm that the main parameter to consider is the WDM total power that co-propagates with the quantum channel; the number of channels and the total bit rate are deduced from the tolerated WDM total power.

Thus, we propose a new figure of merit called CE (for Co-propagation Efficiency) to evaluate the performance of classical & QKD signals co-propagation. The proposed CE metric is given in (1):

$$CE = SKR * P_{WDM} * L \quad (1)$$

It is expressed in Mb/s*mW*km with respectively *SKR* in Mb/s, the Secure Key Rate, $P_{WDM}$, the co-propagating classical channels total power in mW, and L, the link length in km. $P_{WDM}$ represents the total WDM power that co-propagates with the quantum channel. SKR and L refers directly to the ability of the QKD system to deliver a key in a defined configuration. The advantage of this parameter is to be independent from the wavelength of the quantum channel and to incorporate all the factors that are important for practical QKD deployments.

In our case, we obtain a value of CE of 253.7 Mb/s*mW*km in the 50 km configuration (see Fig. 4) with 16.8 dBm aggregated DWDM power which is much higher than in references [10] and [11] for which the CE are respectively 9.3 and 3.18 Mb/s*mW*km. Moreover in [12], a co-propagation over 66 km gives a CE value of 24.9 Mb/s*mW*km. Fig. 4 shows also the very fast degradation of the CE with fiber length and thus directly reflects the ability of a QKD system to tolerate co-propagation of quantum and classical channels.

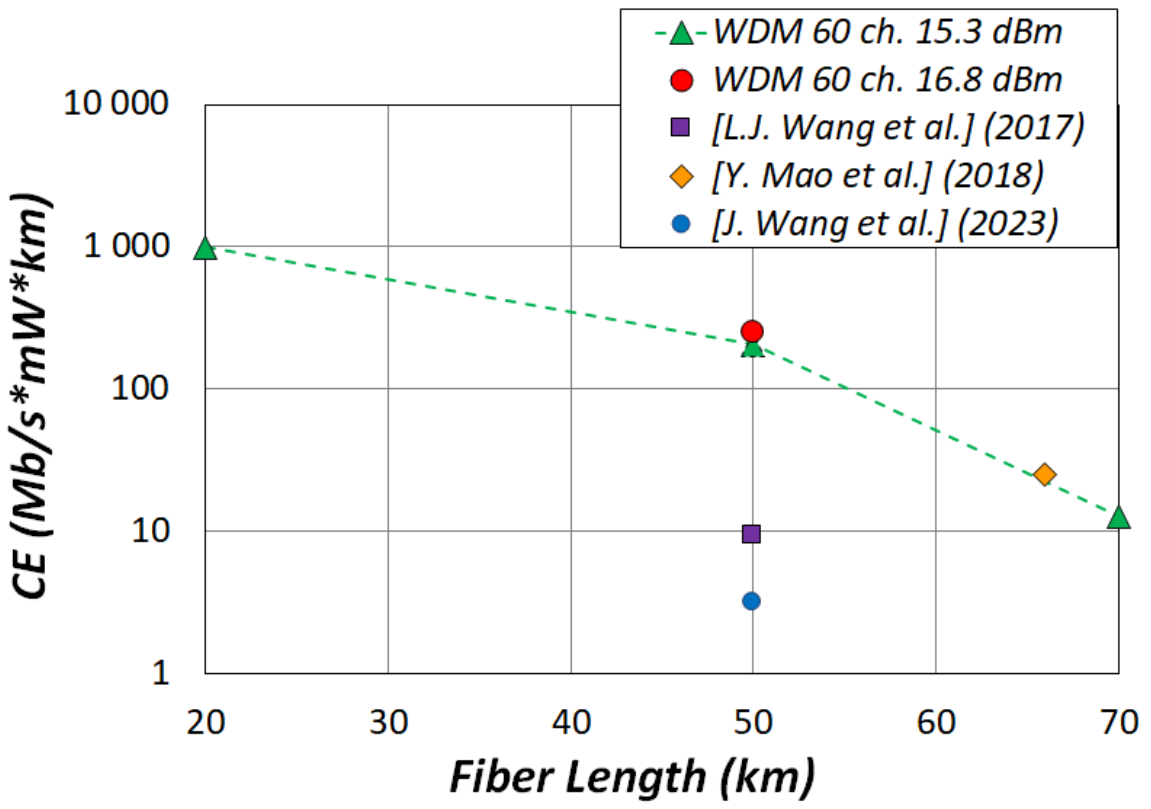

Fig. 4. CE versus fiber length for a constant 15.3 dBm WDM total power, for 16.8 dBm with 50 km and other referenced works.

It's worth mentioning that the comparisons done with this metric (in Fig. 4) have been done only with SSMF experiments. The possibility of generalization to other fibers (like multicore fibers and hollow core ones) is for further studies.

## III. Three Spans Experiment

In the second experiment, whose goal is to validate both this co-propagation possibility in a multi-span chain and demonstrate the encryption of an end to end WDM channel, we use one Multiplexed (MU) and two Long-Distance (LD) QKD systems [13], as shown in Fig 5. The MU-QKD system is the same as the one described in section II. The two LD-QKD systems are optimized for reaching the longest possible transmission distance. This is achieved using a 1550 nm quantum channel as the SSMF losses are minimal in this wavelength range. In this implementation, the quantum channel is transmitted over a dedicated fiber and the three service channels (C59 to C61) (see Fig. 5) are propagating bidirectionally on another fiber on which the WDM channels are also transported. Add/drop multiplexing hardware is also included in the QKD units to multiplex/demultiplex the quantum channel, the service channels of QKD systems (for synchronization, authentication, reconciliation…) and any WDM data channels.

### A. *Description of the optical path*

The transmission experiment consists of a WDM comb of fifty-four DP-QPSK channels at 100 Gb/s ranged from 1533.6 nm to 1557 nm (with 50 GHz spacing) and one coherent 400 Gb/s DP-16-QAM channel at 1542.3 nm carrying the 100 GbE QKD secured signal (as shown in Fig. 6(a)). A spectral hole is created in the WDM comb at 1550-nm by switching-off three channels: this permits the LD-QKD system with its 1550 nm quantum key to work without any disturbing effect. The 400 Gb/s channel operating at ~64 Gbaud has two-fold more power than the 32 Gbaud 100 Gb/s channels. After amplification in EDFA#1 with ~19 dBm output power and ~6 dB noise figure, the mixed 100G/400G WDM comb is sent to the auxiliary (AUX) Rx port of the Alice#1 terminal of the first LD-QKD system. In that case the WDM comb aggregated power in the fiber at the QKD classical (C) port is ~17.5 dBm, due to an insertion loss of ~1.5 dB on the path between the AUX Rx and QKD C ports (see Fig. 5). From the QKD C port of Alice#1 come out the WDM comb as well as two QKD service channels (C59 & C60: blue arrows), while the QKD quantum (Q) port of Alice#1 delivers the quantum signals (green arrow). Two separated fibers of 67 km length propagate the quantum channel and the 1550 nm range signals (service and WDM channels), respectively. Another QKD service channel (C61) emitted by the QKD C port of Bob#1 counter-propagates with the WDM comb and C59 & C60 service channels on the same fiber, as shown in Fig. 5. At the Bob#1 side terminal, the quantum signal is recovered on the Q port of the QKD unit while the classical channels (i.e. WDM, C59 & C60) are recovered by the C port. The demultiplexing unit of Bob separates the WDM comb from the service channels and sends the WDM signals to the AUX Tx port. The WDM comb is re-amplified with a ~19 dBm EDFA#2 and sent to the AUX Rx port of the Alice#2 terminal of the MU-QKD system. The operation described above for the LD-QKD system is the same for the MU-QKD units, except that the QKD service channels C59 & C60, the WDM comb and the quantum channel at 1310 nm are coupled all together and come out from the QKD Tx port of Alice#2 on a same SSMF section of 50 km. A second 50 km SSMF span is employed to propagate the backward service channel C61. At the Bob#2 terminal, the demultiplexing unit separates the quantum channel from the service channels and WDM comb (sent to the AUX Tx port). The WDM comb is re-amplified with a ~19 dBm EDFA#3 and sent to the AUX Rx port of the Alice#3 terminal of the second LD-QKD system, the last of the chain. The operation inside this last QKD unit is fully identical to the first one. At the end of the link, whose spectrum is depicted in Fig. 6(b), a flat-top tunable wavelength / bandwidth optical filter is used to extract the 400 Gb/s channel from the received WDM comb. Note that the experiment performed here is a single direction one but, in a real network implementation, the WDM traffic would have to propagate from the right end of the link to the left end in order to create a bidirectional end to end link. In that case, we should add a third fiber between Alice#1 and Bob#1 and another between Alice#3 and Bob#3. But thanks to the co-propagation of the quantum channel with the WDM channels in the MU system we would not need any other fiber between Alice#2 and Bob#2, which is a huge advantage. Moreover, very often in the network, the fibers are working in pairs, so the addition of a third fiber would often lead to the reservation of a pair even if we don't use the fourth fiber.

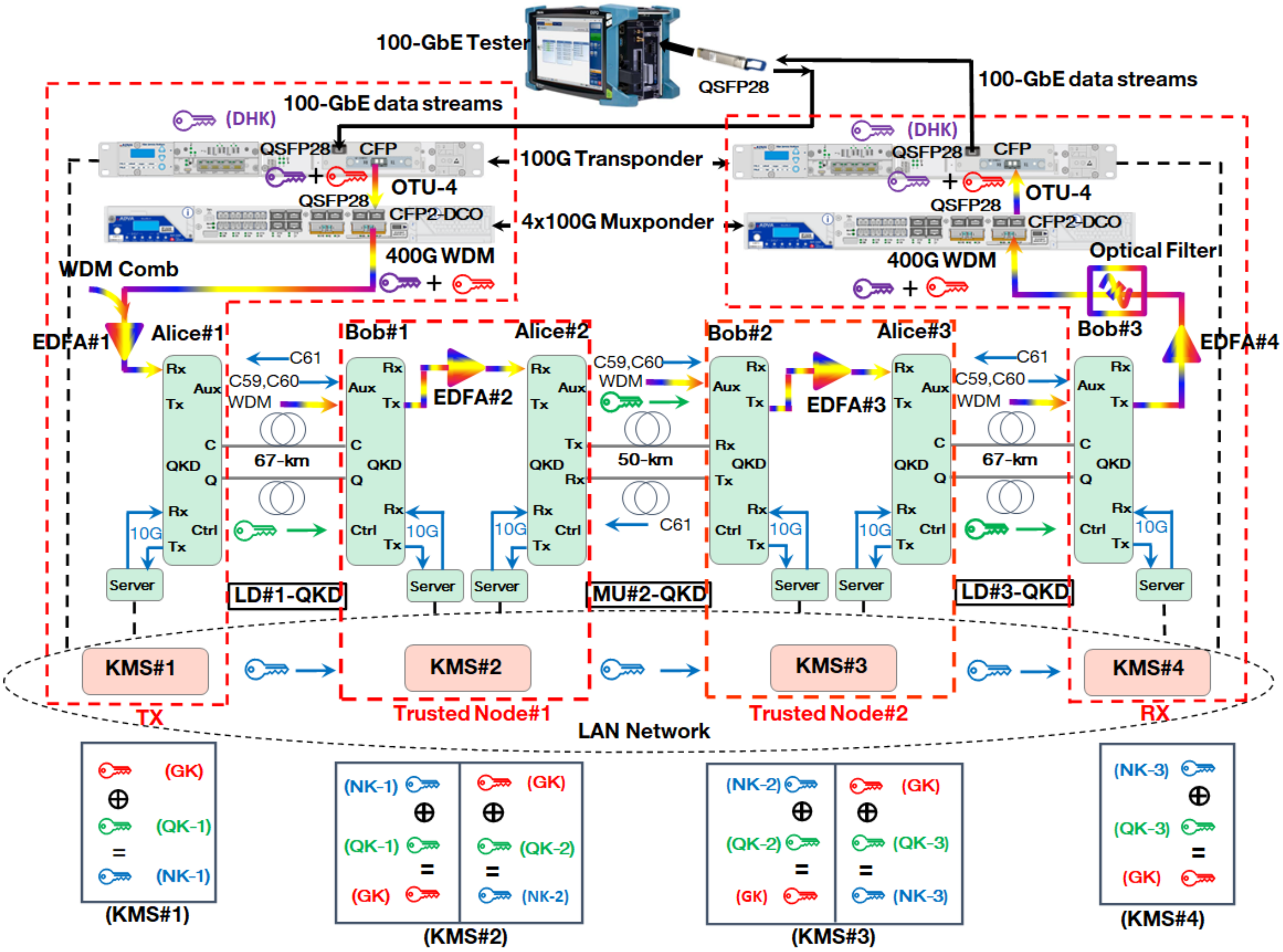

Fig.5. Experimental set-up with the three QKD links, the 100 GbE tester, the 100G transponders, the Nx100G muxponders (with N=4), the KMSs, the two trusted nodes, and the description of key management by all the elements of the set-up.

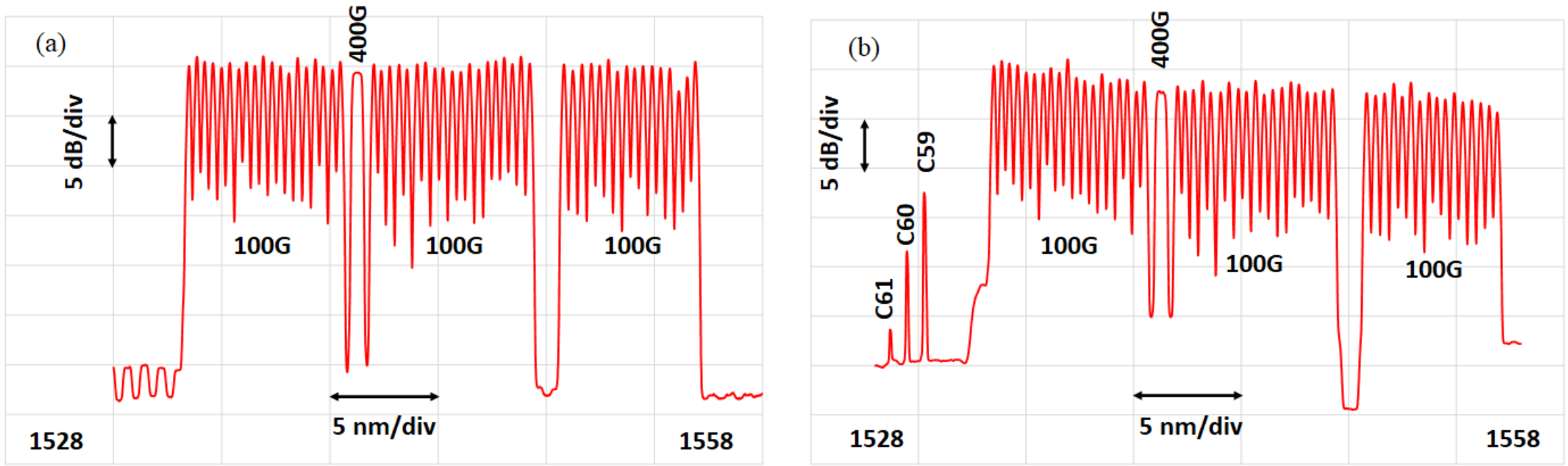

Fig.6. (a) WDM comb sent into the transmission line, (b) WDM comb recovered at the Rx side.

*B. Description of the encryption process*

A 100 GbE tester generates the data flows that are converted in the optical domain by a client or "grey" 100GBASE-LR4 QSFP28 optic [14]. A second 100GBASE-LR4 QSFP28 interface plugged inside a 100G transponder [15] receives the data. This transponder is able to map and encrypt the 100 GbE data stream as OTNsec [5] thanks to the Advanced Encryption Standard AES-256 algorithm. Complying with the NIST recommendation [16], the effective 256-bit cipher key is derived from an embedded Diffie-Hellman Key (DHK) (in purple) exchange and a QKD "Global Key" (GK) (in red) delivered by a Key Management Server (KMS) [17], as shown in Fig. 5. At the other side of the transponder, a WDM 100G CFP Digital Coherent Optic (DCO) that colors the signal, or a 100GBASE-LR4-CFP transceiver that keeps the signal in the "grey" domain can be used. This latter configuration is chosen here, due to only a short back-to-back connection between the 100G transponders and 4x100G muxponders. The 100 GbE data stream is then mapped and encrypted in an OTU-4 [18] container before being delivered by the 100G transponder. It is then detected by a third 100GBASE-LR4 QSFP28 optic (also operating in the OTU-4 mode) plugged inside a Nx100G muxponder [15] (with N=4) that mixes the 100 GbE QKD secured signal with other three 100 GbE non-encrypted data flows. At the output of the muxponder, a coherent 400 Gb/s DP-16QAM signal is generated thanks to a 400G CFP2-DCO [14]. The role of the 100G transponder is thus to encrypt the data with a "secret key" derived from the QKD GK provided by the KMS#1 and a DHK, while the function of the muxponder is to encapsulate the 100G QKD-secured signal in a 400 Gb/s WDM pipe. Note that the reverse process is performed at the receiver side to decrypt and recover the 100 GbE data in the 100 GbE tester.

*C. Key management*

The Q-KMS, installed on a server at each node, handles the keys inside the system and establishes the link between all the elements of the set-up, all of them being connected to a Local Area Network (LAN), as shown in Fig. 5. To establish a trusted chain between the different devices of the experiment, a "root" Certification Authority (CA) is used to create certificates to an "intermediate" CA that itself issues certificates to the end entities (i.e. the equipment of the set-up) so that each element of the experiment trusts their CA and certificates issued by their CA. This constitutes a Public Key Infrastructure (PKI) as per X.509v3 [19]. Once this certification is carried out, the key exchange between the various elements of the experiment can start. Note that certificates are used for Transport Layer Security (TLS) protection of key delivery from KMS to encryptors only (as per the ETSI QKD 014 Key Delivery API international standard), within the trusted perimeter of the operator's building. Such PKI is known to be vulnerable to the quantum computing threat but is considered acceptably safe within an operator's secure physical perimeter. All data travelling between buildings (i.e. over deployed fiber) are QKD-secured and do not rely on PKI.

The paired 100G transponders receive 256-bit "secret keys" (i.e. red keys → GK, in Fig. 5) from KMS#1 and KMS#4, for the transmitter and receiver, respectively. The transponder mixes one of these 256-bit GK with a 512-bit DHK (purple) generated inside the 100G transponder through a Key Derivation Function (KDF) [16]. The resulting key (i.e., DHK + GK) is used to encrypt the 100G OTU-4 data streams. This process of refreshing the key session is performed every 60 seconds. In order to provide an end to end key transfer, the KMS#1 executes a XOR operation between GK and a first "quantum key" (in green → QK-1) to obtain a "network key" (in blue → NK-1) that can be considered as "public" and transmitted through the LAN up to the KMS#2, part of the first trusted node. The KMS#2 performs a XOR operation between NK-1 and QK-1 (transmitted from Alice#1 to Bob#1) so that GK can be recovered. The KMS#2 carries out a XOR operation between GK and a new quantum key (QK-2) so that to obtain a new network key (NK-2) sent to the KMS#3. The operation is repeated until reaching the KMS#4 where GK is extracted and sent to the end link 100G transponder where the 100G OTU-4 data stream is deciphered by a process reverse to the one described above.

*D. Results analysis*

Two kinds of tests were performed during the evaluation. First, we measured the ability of the whole QKD chain to provide ciphering keys at both ends of the link. Fig. 7 shows the information via the ETSI GS 014 interface on the Network Management System (NMS) of the 100G transponder. Each encryption service channel requests a new QKD key every minute from the key pool of the QKD system. The instant key count in the key pool can be requested using the "Get Status" API as per ETSI GS 014. In our setup, the QKD Tx (and Rx) is configured by the KMS to store a maximum of 1000 keys (of 256 bits) for AES encryption use. In Fig. 7, we can see the successful operation of the key management process in the fact that the ″stored key count″ at Tx side has decreased from 1000 to 823 (see the dotted line rectangle), thus demonstrating that the 100G tributary of the 4x100G muxponder correctly uses the quantum keys.

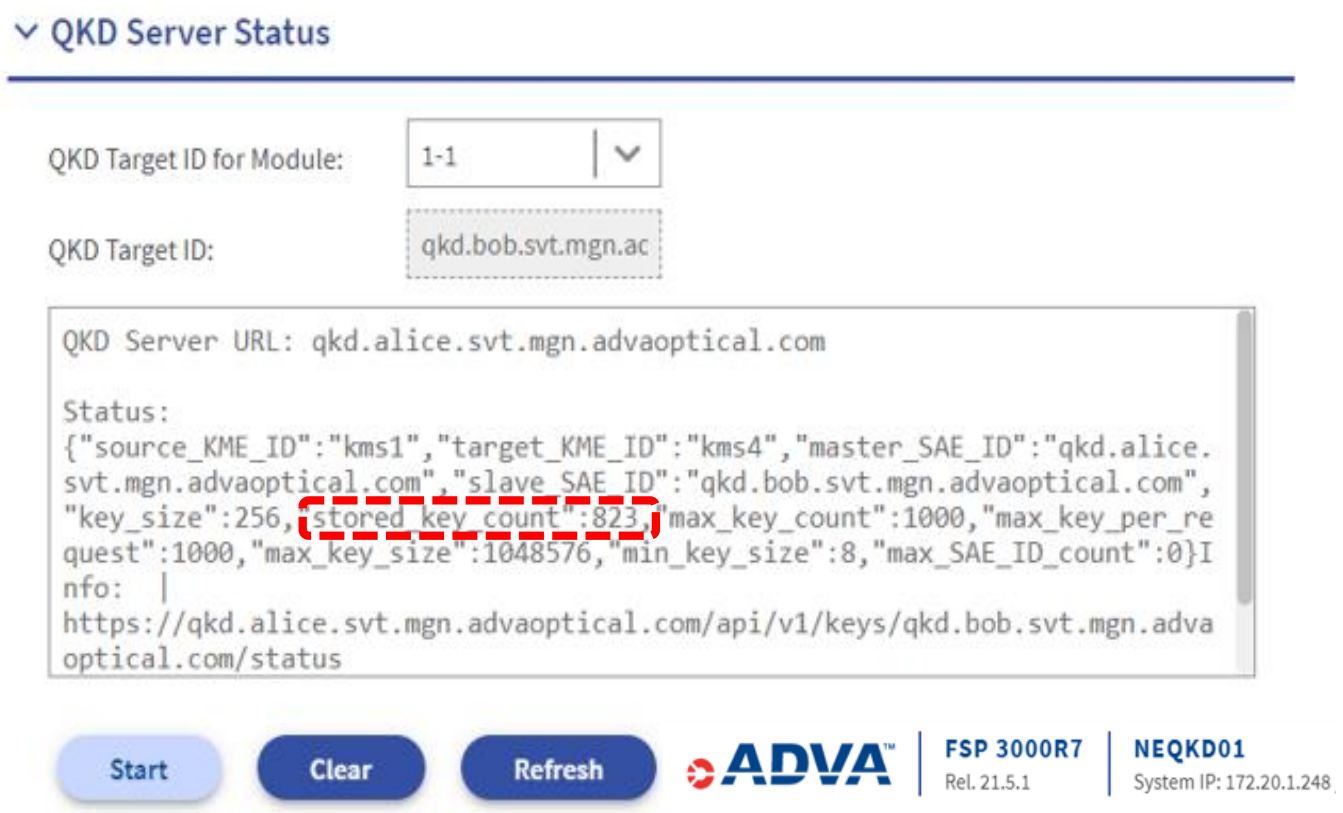

Fig.7. QKD server status recovered on the NMS of the 100G transponder at the Tx side with all the information related to the key management.

The second tests concerned the available SKRs and the QBERs for the three QKD systems. Various levels of the WDM comb total power injected into the SSMF have been evaluated. They were ~17.5, ~15.5 and ~13.5 dBm. As expected, the mean SKR over 24 hours of the two LD-QKD systems is higher than the one of the MU-QKD system (i.e., ~260 and ~272 kb/s against ~36.5 kb/s at ~17.5 dBm, ~48 kb/s at ~15.5-dBm and ~63 kb/s at ~13.5 dBm). The end to end SKR is inferred from the lowest SKR, namely the one of the MU-QKD system. Identically, the mean QBER is better for the LD-QKD systems than for the MU-QKD one (i.e., ~3% and 3.2% against ~6.7% at ~ 17.5 dBm, ~6% at ~15.5 dBm and 5.4% at ~13.5 dBm).

Finally, we also considered the pre-Forward Error Correction code (pre-FEC) BER of the 400G channel, for the various levels of the WDM comb power injected into the SSMF. Note firstly that the pre-FEC BER limit of the OFEC [9] (used in the 400G CFP2-DCO) was measured at ~$1.75 \times 10^{-2}$ and corresponds to a Required Optical Signal to Noise Ratio (ROSNR) in Back to Back (BtB) of ~22.2 dB in 0.1 nm. In all cases, the pre-FEC BER of the 400G channel is well below the OFEC BER threshold. The pre-FEC BER as well as the received OSNR measured in 0.1 nm are respectively equal to ~$3 \times 10^{-3}$ and ~29.7 dB at ~17.5 dBm, ~$5.5 \times 10^{-3}$ and ~27.7 dB at ~15.5 dBm, ~$1 \times 10^{-2}$ and ~26.2 dB at ~13.5 dBm, namely largely above the ROSNR in BtB, that ensures error-free transmission of 400G and 100 GbE data flows at the end of the chain. More results can be found in [20].

## IV. Conclusion

We have first reported experimental results of the co-propagation of the quantum channel of a QKD system at 1310 nm with a DWDM comb of 60 channels at 100 Gb/s for a total rate of 6 Tb/s in a single span experiment. The emission of a secure key is possible with a very high aggregated power: ~17 dBm WDM total power (limited by the set-up) for 50 km SSMF and ~16 dBm for 70 km.

We have proposed a new figure of merit to evaluate the co-propagation efficiency and, based on this metric, the results presented here outperforms the previous co-propagating results in close conditions.

We have also demonstrated the encryption of a 100G interface of a coherent 400 Gb/s DP-16QAM channel that transport a QKD-secured 100 GbE data stream over 184 km of SSMF through three QKD systems and two in-line trusted nodes with other fifty-four DWDM channels at 100 Gb/s. One of the QKD section co-propagates over 50 km the quantum key and the WDM comb of ~17.5 dBm aggregated power, namely a power level perfectly compliant with the modern WDM systems deployed in the field.

To conclude, for the widespread deployment of commercial QKD technology into existing telecommunication networks, it is essential that the quantum channel can co-propagate with existing classical data channels on the same fiber. This is a prerequisite for many Data Center Interconnection (DCI) and metropolitan applications, which comprise existing fully filled WDM links with 100 Gb/s and 400 Gb/s channels. Our results demonstrate how current QKD systems can be integrated into telecom operator networks to meet these requirements, showing both record co-propagation performance and end to end quantum-secure communications over a multi-span link. This paves the way to larger scale quantum-secure networks.

## References

[1] S. Pirandola et al, "Advances in quantum cryptography," *Adv. Opt. Photon.*, vol.12, pp. 1012-1236, 2020, DOI: 10.1364/AOP.361502

[2] R. Kumar, H. Qin and R. Alléaume, "Coexistence of continuous variable QKD with intense DWDM classical channels" *New J. Phys.* vol.17, 2015, Art. no. 043027, 2015, DOI: 10.1088/1367-2630/17/4/043027

[3] F. Honz et al., "First demonstration of 25λ × 10 Gb/s C+L band classical / DV-QKD co-existence over single bidirectional fiber link," *J. Lightw. Technol.*, vol. 41, no. 11, pp. 3587-3593, Jun. 2023, DOI: 10.1109/JLT.2023.3256352

[4] P. Gavignet, F. Mondain, E. Pincemin, A. Grant, L. Johnson, R. Woodward, J. Dynes, and A. Shields, "Co-propagation of 6 Tb/s (60*100Gb/s) DWDM & QKD channels with ~17 dBm aggregated WDM power over 50 km standard single mode fiber," in *Proc. Opt. Fiber Commun. Conf.*, 2023, Paper Tu3H.2, https://arxiv.org/abs/2305.13742

[5] "OTN Sec : Security for OTN beyond 100 Gbps", ITU-T SG15, 2013, https://www.itu.int/md/T13-SG15-C-0110/en

[6] M. Lucamarini et al., "Efficient decoy-state quantum key distribution with quantified security," *Opt. Exp.*, vol. 21, pp. 24550-24565, 2013.

[7] Z. Yuan et al., "10-Mb/s Quantum Key Distribution," *J. Lightw. Technol.*, vol. 36, no. 16, pp. 3427-3433, Aug. 2018.

[8] E. Pincemin et al., "Interoperable CFP-DCO and CFP2-DCO Pluggable Optic Interfaces for 100G WDM Transmission," in *Proc. Opt. Fiber Commun. Conf.*, 2019, Paper Th1I.3, https://doi.org/10.1364/OFC.2019.Th1I.3

[9] E. Pincemin et al., "927-km End-to-End Interoperable 400-GbEthernet Optical Communications through 2-km 400GBASE-FR4, 8×100-km 400G-OpenROADM and 125-km 400-ZR Fiber Lines," in *Proc. Opt. Fiber Commun. Conf.*, 2022, Paper Th4A.3, https://doi.org/10.1364/OFC.2022.Th4A.3

[10] L.J. Wang et al., "Long-distance copropagation of quantum key distribution and terabit classical optical data channels," *Phys. Rev. A*, vol. 95, 2017, Art. no. 012301.

[11] J. Wang et al., "Time-Interleaved C-band Co-Propagation of Quantum and Classical Channels", 2023, https://arxiv.org/abs/2304.13828v2

[12] Y. Mao, "Integrating quantum key distribution with classical communications in backbone fiber network", *Opt. Exp.*, vol. 26, no. 5, pp. 6010-6020, 2018, https://doi.org/10.1364/OE.26.006010

[13] "MU-QKD and LD QKD systems specifications", 2023. https://www.toshiba.eu/quantum/products/quantum-key-distribution/

[14] "Physical layers and management parameters for 100 Gb/s and 400 Gb/s operation over single mode fiber at 100 Gb/s per wavelength", IEEE/ISO/IEC 8802-3:2021/Amd 11-2021, 2022. https://standards.ieee.org/ieee/8802-3_2021_Amd_11/10988/

[15] "FSP-3000 Transponders and Muxponders specifications", Adtran, 2023, https://www.adtran.com/en/products-and-services/open-optical-networking/fsp-3000-open-terminals/transponders-and-muxponders

[16] "Recommendation for Key-Derivation Methods in Key-Establishment Schemes," 2020, https://csrc.nist.gov/publications/detail/sp/800-56c/rev-2/final

[17] ETSI GS QKD 014 v.1.1.1, "QKD; Protocol and data format of REST-based key delivery API", 2019, https://www.etsi.org/deliver/etsi_gs/QKD/001_099/014/01.01.01_60/gs_QKD014v010101p.pdf

[18] "OTU-4 Long-Reach Interface Recommendation", ITU-T Rec. G.709.2/Y.1331.2, ITU, Geneva, Switzerland, 2018. https://www.itu.int/ITU-T/recommendations/rec.aspx?rec=13522

[19] IETF, "Internet X.509 Public Key Infrastructure Certificate and Certificate Revocation List (CRL) Profile," RFC 5280, 2008. https://datatracker.ietf.org/doc/html/rfc5280

[20] E. Pincemin et al., "400G Transmission of QKD-Secured 100G Data Stream over 184 km SSMF through three QKD Links and two Trusted Nodes", in *Proc. Eur. Conf. Opt. Commun.*, 2023, Paper M.A.4.4.